\documentclass[
 ,final            
  ]
  {aipproc}

\layoutstyle{8x11double}

\usepackage{color}

\def\lsim{\:\raisebox{-0.5ex}{$\stackrel{\textstyle<}{\sim}$}\:}
\def\gsim{\:\raisebox{-0.5ex}{$\stackrel{\textstyle>}{\sim}$}\:}

\newcommand{\tanb} {\tan\beta}
\newcommand{\mhpm} {m_{H^{\pm}}}
\newcommand{\mhone} {m_{h_{1}}}
\newcommand{\hone} {h_{1}}
\newcommand{\htwo} {h_{2}}
\newcommand{\mhtwo} {m_{h_{2}}}

\newcommand{\ra} {\rightarrow}
\newcommand{\nbtag}  {N_{btag}}
\def\met        {E\!\!\!\!/_T}

\newcommand{\beq} {\begin{equation}}
\newcommand{\eeq} {\end{equation}}
\newcommand{\bea}{\begin{eqnarray}}
\newcommand{\eea}{\end{eqnarray}}

\begin{document}

\title{CP-violating Higgs at Tevatron}


\author{Siba Prasad Das}{
 address={Physikalisches Institut der Universit\"at Bonn,
Nu\ss allee 12, D-53115 Bonn, Germany } 
}

\author{Amitava Datta}{
  address={Indian Institutes of Science Education and Research, 
Salt Lake City, Kolkata - 700106, India }
}

\author{Manuel Drees}{
  address={Physikalisches Institut der Universit\"at Bonn,
 Nu\ss allee 12, D-53115 Bonn, Germany }
}

\begin{abstract}
  
  We analyze the prospect for observing the intermediate neutral Higgs
  ($h_2$) boson in its decay to two lighter Higgs bosons ($h_1$) at the
  Tevatron in the framework of the CP violating MSSM using the PYTHIA event
  generator. We consider the lepton+ 4-jets+ $\met$ channel from $p \bar p \ra
  W h_2 \ra W h_1 h_1 \ra l \nu_l b \bar b b\bar b$, with two or three tagged
  $b$ jets. We found that it is very hard to observe this signature in the
  LEP-allowed region of parameter space, due to the small signal efficiency.

\end{abstract}

\maketitle


\section{Introduction}

The Minimal Supersymmetric Standard Model (MSSM) requires two Higgs doublets,
leading to a total of five physical Higgs bosons, two neutral CP-even, one
neutral CP-odd and two charged. In the presence of CP violation, the two
CP-even ($\phi_1$ and $\phi_2$) and one CP-odd ($a$) eigenstates can mix
radiatively \cite{CPmixing0,Lee:2003nt}. The mass eigenstates $h_1$, $h_2$ and
$h_3$ with $m_{h_1} < m_{h_2} < m_{h_3}$ can be obtained from the interaction
eigenstates $\phi_1$, $\phi_2$ and $a$ with the help of the orthogonal matrix
$O_{\alpha i}$, $(\phi_1,\phi_2,a)^T_\alpha= {O_{\alpha i}}(h_1,h_2,h_3)^T_i
\, ,$ which diagonalizes the Higgs boson mass matrix. $O$ depends on various
parameters of the SUSY Lagrangian.

Due to this mixing, the Higgs mass eigenstates no longer are CP eigenstates.
Moreover, the masses of the Higgs bosons, their couplings to SM and MSSM
particles, and their decays are significantly modified \cite{Lee:2003nt}. For
example, the Higgs boson couplings to pairs of gauge bosons is scaled by
$g_{_{h_iVV}}$ relative to the SM. These couplings can be expressed as
${{g_{_{h_iVV}} }} = \cos\beta\, O_{\phi_1 i}\: +\: \sin\beta\, O_{\phi_2 i}
\, ,$ where $\tan\beta$ is the ratio of Higgs VEVs. The magnitude of  
$g_{_{h_2 W W}}$  is directly related to the production process  
studied in this paper.

In our numerical analysis, we chose the $CPX_{0.5}$
scenario with maximal CP violation~\cite{Carena:2000ks},  
\begin{eqnarray}
\widetilde{M}_Q &=& \widetilde{M}_t =
\widetilde{M}_b = 500 \ {\rm GeV},\qquad
\mu  = 4 \widetilde{M}_Q \,,\nonumber\\
|A_t| &=& |A_b| = 2 \widetilde{M}_Q,\qquad
{\rm arg}(A_t) = {arg}(A_b)  =  90^\circ\,, \nonumber\\
|m_{\tilde{g}}| &=& 1~{\rm TeV}\,,\qquad
{\rm arg}(m_{\tilde{g}})\ =\ 90^\circ\, .
\label{eq:CPX}
\end{eqnarray}
The remaining two input parameters are the charged Higgs boson mass $\mhpm$
and $\tanb$. We calculated the spectrum and the couplings using {\tt
CPsuperH}~\cite{Lee:2003nt}. 

It is quite well known that the LEP experiment were not able to exclude
certain regions in the $\mhone-\tan\beta$ plane, where $h_1$ is dominantly a
CP-odd state with almost vanishing coupling to the gauge bosons while $h_2$ is
just too heavy to be produced. One region has $M_{h_1}\lsim 10$ GeV, so that
$\hone \ra \tau^+ \tau^-$ is dominant; in the other, $M_{h_1} \sim 30 - 50$
GeV so that $\hone \ra b \bar b$ is dominant. These occur for $\tan\beta$ in
between 3-10~\cite{LEP_allow}. We analyze the prospect for observing the
intermediate neutral Higgs ($\htwo$) in the second of these LEP allowed 
regions.

\smallskip

\section{Numerical Analysis}

In our simulation we used the {\tt PYTHIA v6.408 } \cite{Sjostrand:2006za}
event generator with the {\tt SLHA} \cite{Skands:2003cj} input at Tevatron
Run-II with $\sqrt s =1.96$ TeV. We used {\tt MadGraph/MadEvent v4.2.8}
\cite{Maltoni:2002qb} for generating parton level SM backgrounds which were
fed to PYTHIA for showering. We set the renormalization and factorization
scale to $Q= \sqrt {\hat s}$ and used CTEQ5L for the parton distribution
functions (PDF).

The signal arises from $p \bar p \rightarrow W \htwo \rightarrow \ell
\nu_{\ell} \hone\hone \rightarrow \ell \nu_{\ell} b \bar b b \bar b$, leading
to $\ell jjjj \met$ events, where $\ell = e$ or $\mu$. The effective
cross section for this signal topology can be expressed as, 
\bea  \label{effcross}
C_{211_{4b}} & = & \sigma_{SM}(p \bar p \rightarrow W \htwo) {g^2_{_{h_2 W W}}} 
Br(\htwo \rightarrow \hone\hone) \nonumber \\ && 
\times {Br(\hone \rightarrow b \bar b)}^2 2{Br(W \rightarrow e \nu_{e})}\,,
\eea
where $W$ stands for $W^{\pm}$ and the factor 2 is for $\rm \ell = e ~and
~\mu$. 

We have used the toy calorimeter simulation ({\tt PYCELL}) provided in {\tt
  PYTHIA} with the following criteria: calorimeter coverage is $\rm |\eta| <
3.6$; the segmentation is given by $\Delta \eta \times \Delta \phi$=$0.16
\times 0.098$ which resembles the CDF detector; Gaussian smearing of the total
energy of jets and leptons; a cone algorithm with $\rm\Delta R(j,j)=0.4$ has
been used for jet finding; $\rm E_{T,min}^{cell} \ge 1.5$ GeV is considered to
be a potential candidate for jet initiator; minimum summed $\rm
E_{T,min}^{jet} \ge 7.0$ GeV is accepted as a jet and the jets are ordered in
$E_{T}$; leptons ($\rm \ell = e, ~\mu$) are selected with $\rm E_T \ge 15.0$
GeV and $\rm |\eta| \le 2.0$ and no jet should match with a hard lepton in the
event.
       
A jet with $\rm |\eta| \le 1.2$ matched with a $b-$flavored hadron $B$, i.e.
with $\Delta R(j, B) < 0.2$, is considered to be {\em taggable}. We have
treated $b$ tagging in these taggable jets taking into account the $E_T$
dependent tagging probability, following Fig.6 (top) of \cite{Hanagaki:2005fz}.
We find that our tagging algorithm agrees well with the $t \bar t$
analysis of CDF \cite{cdfttbar}.


We chose $\mhpm$= 132 GeV and $\tanb= 4.0$ with $CPX_{0.5}$ as a benchmark
point denoted by CPX-1, for which the effective cross-section
(Eq.\ref{effcross}) is maximal. The masses of the Higgs bosons, $\mhone$ and
$\mhtwo$, are 36.0 GeV and 101.6 GeV respectively. $\sigma_{SM}$ and
$C_{211_{4b}}$ are 217 fb and 21.6 fb respectively.

We have applied the following selection cuts: 
\begin{itemize}
\item S1: $N_{\rm jet} \ge 4$ ; 
\item S2: $E_{T}^{j1,j2,j3,j4} > 10.0$ GeV and $ |\eta^{j1,j2,j3,j4}| < 3.0$;
\item S3: $N_{\rm lepton} \ge 1$, $E_{T}^{l} > 15.0$ GeV and $|\eta^{l}| <
  2.0$; 
\item S4: $\met > 15$ GeV where $\met$ is calculated from all visible
  particles;  
\item {S5a(b): $\nbtag \ge 2~(3)$}; $|\eta^{b-jet}| < 1.2$, $\Delta R(j,B)\le
  0.2$;  
\item S6($6'$): $H_{T} = \met + \sum_{\rm obj} E_{T} \le 300~(250) \ {\rm
    GeV}$;
\item S7($7'$): $\Delta \phi(b1,b2) \le 2.09 (1.57) $;
\item S8($8'$): $N_{\rm obj}= N_{\rm lepton} +  N_{\rm jet} \le 6~(5)$. 
\end{itemize}
\begin{table}
\begin{tabular}{lrrrrrr}
\hline
  & \tablehead{1}{r}{b}{C1-4 }
  & \tablehead{1}{r}{b}{I5a}
  & \tablehead{1}{r}{b}{I5b}
  & \tablehead{1}{r}{b}{I6}
  & \tablehead{1}{r}{b}{I7} \\
\hline
CPX-1 & .155 & .102  &.009  & .937 &.810 \\
\hline
$t \bar t$ & .511 & .108 & .00021 & .462 &  .481   \\
$W b \bar b$ &.014 & .015 & .00002 &  .986  &  .502  \\
ZZ & .026 & .118 & .01735 &     .945 & .434  \\
$WZ$ & .054 & .048 & .00008 &  .968 &  .432   \\
$W t \bar t$ &.619 & .127  & .00062 & .102 & .530 \\
\end{tabular}
\caption{Individual selection efficiencies for signal and major backgrounds at
  Tevatron Run-II. C1-4 stands for the combined efficiencies due to S1, S2, S3
  and S4 selections. Mistagging has not been included here.} 
\label{tab:EffTeV}
\end{table} 

\begin{figure}
\includegraphics[height=.24\textheight,width=0.48\textwidth]{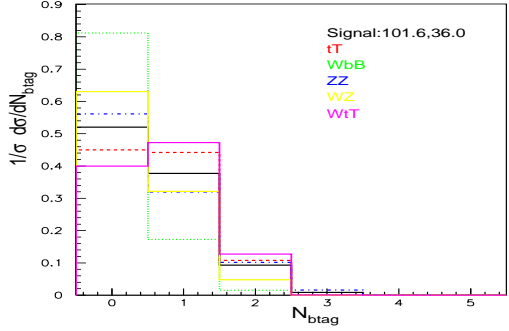}
\caption{{\small Normalized distribution of the number of $b$ tags at Tevatron
    Run-II for the signal and the backgrounds without inclusion of
    mis-tagging.}}
\label{fig:nbtagTeV}
\end{figure}

We have displayed the cumulative (C) and individual (I) selection efficiencies
for the signal and the major backgrounds in Table.~\ref{tab:EffTeV}. Not
surprisingly, $t \bar t$ is the main source of background after applying the
basic cuts S1 to S4. The $\nbtag$ distribution is shown in
Fig.~\ref{fig:nbtagTeV}. The individual $\nbtag \geq 2$ efficiencies (I5a) are
almost the same for signal and $t \bar t$, see Table~\ref{tab:EffTeV}. The
signal contains more $b$ quarks, but the $t \bar t$ background has much harder
$b$ jets, leading to larger tagging probabilities. However, this background
can contain a third $b$ jet only due to showering. Hence requiring $\nbtag
\geq 3$ reduces the individual signal ($t \bar t$) efficiency by
$\mathcal{O}(10^{-1(-3)})$; we thus expected that requiring $\nbtag \geq 3$
might be useful. In order to check this, we have to allow for mis-tagging
non-$b$ jets as $b$ jets. We assumed a mis-tagging probability of 1\% for
$u,d,s$ and gluon jets, and 10\% for $c$ jets \cite{Hanagaki:2005fz,
  Bowen:2004my}. We then re-evaluated I5a (I5b) only for $t \bar t$ and the
efficiency increases from 10.8\% to 12.5\% (0.02\% to 0.5\%). 

\begin{figure}
\includegraphics[height=.24\textheight,width=0.48\textwidth]{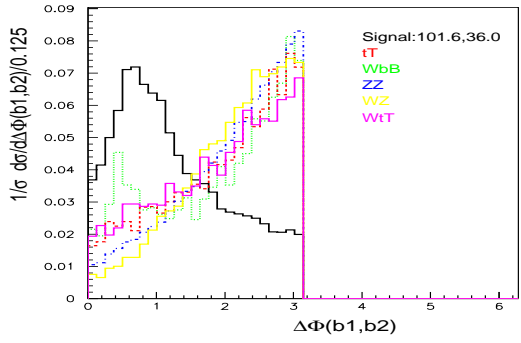} 
\caption{{\small Normalized distribution of the azimuthal angle between the
    first two $b$-tagged jets at Tevatron Run-II for the signal and the
    backgrounds without inclusion of mis-tagging.}}
\label{fig:deltaphi}
\end{figure}

We also found that the azimuthal angle of the first two $b$-tagged jets,
$\Delta\phi(b1,b2)$, for the signal (dominant backgrounds) has its maximum
around 0.7 (2.8) (see Fig.\ref{fig:deltaphi}). Clearly proper choices of
$\nbtag$ and $\Delta\phi(b1,b2)$ could be potential discriminators to isolate
the signal from backgrounds. We have also checked that upper cuts (vetos) on
$H_{T}$ and $N_{\rm obj}$ suppress the higher mass and higher multiplicity,
for example, $W t \bar t$ and $t \bar t$ events. 

\begin{figure}
\includegraphics[height=.24\textheight,width=0.48\textwidth]{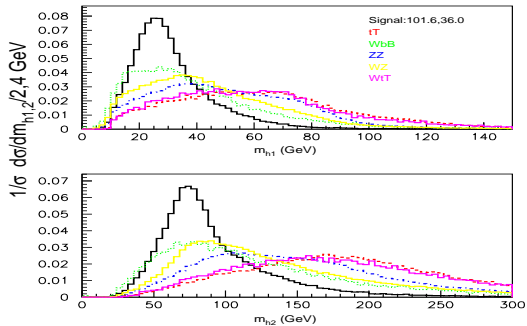}
\caption{{\small Normalized distributions for $M_{j_1j_2}$ $\approx$ $M_{j_3j_4}$ (top)    
    and $M_{j_1j_2j_3j_4}$ (bottom) at Tevatron Run-II for the signal
    and the backgrounds. }}
\label{fig:hmass}
\end{figure}

Furthermore, the veto on $N_{\rm obj}$ facilitates reconstructing the mass of
the Higgs bosons, $\mhone$ and $\mhtwo$, by reducing the combinatorics. To
that end, we calculated all possible di-jet invariant masses ($M_{jj}$) for
events with $N_{jet} \ge 4$; pairs of jets for reconstructing each $\mhone$
were selected by minimizing $|M_{{j_{1}}{j_{2}}} - M_{{j_{3}}{j_{4}}}|$, with
$M_{{j_{1}}{j_{2}}} + M_{{j_{3}}{j_{4}}} \geq 20$ GeV. The reconstruction of
$\mhtwo$ is then straightforward, {\it i.e.}, $m_{h_2} = M_{j_1j_2j_3j_4}$,
see Fig.\ref{fig:hmass}.

We find that for our benchmark point the number of signal ($t \bar t$ with
$\sigma_{t \bar t}= 5$ pb) events surviving after the cumulative selections
(C1-8) is 3 (135) for 10 fb$^{-1}$ of integrated luminosity ($\int{\cal L} dt$)
with double $b$ tagging; the total number of background events is
approximately 165. Thus the significance ($ S \over \sqrt(B)$) is $\approx $
0.23. Requiring at least three $b$ tags leaves only about 1 signal event, even
before applying the remaining cuts. Similarly, requiring only two $b$ tags but
applying the more stringent cuts $6'+7'+8'$, the number of surviving signal
($t \bar t$) events $\approx$ 1 (12). We conclude that this signal is
impossible to observe at Tevatron Run-II.

%


We have extended this study to the LHC \cite{atlas} with the following
pre-selection cuts: $E_{T}^{j} > 25$ GeV, $|\eta^{j}| < 3.0$; $E_{T}^{l} > 20$
GeV and $ |\eta^{l}| < 2.5 $; taggable $b$ jets require $|\eta^{b-{\rm jet}}|
< 2.5$, $\Delta R(j,B) \le 0.2$ with $\epsilon_b=50$\%. We assumed the same
mis-tagging probability as at the Tevatron \cite{Hanagaki:2005fz,
  Bowen:2004my}. The effective cross-section ($C_{211_{4b}}$) for the Signal
at LHC is 390.1 fb. Keeping the Tevatron selection criteria, we find that the
individual signal efficiency for $\nbtag \gsim $ 2 (3) is 15.0\% (1.76\%). The
corresponding efficiency for $t \bar t$ is 18.6\% (1.0\%). The number of
surviving signal ($t \bar t$ with $\sigma_{t \bar t}$= 5$\times 10^5$ fb)
events with triple $b$ tag after the cumulative cuts C1-8, is $\approx$ 10
(3088) for $\int{\cal L} dt$=100 $fb^{-1}$; using instead the more stringent
cuts $6'+7'+8'$, these numbers reduce to $\approx$ 1 (687). Although these
cuts have not yet been optimized for the LHC, these results are not very
promising, either.

\section{conclusions}

We analyzed the possibility of observing neutral Higgs bosons at the Tevatron
in the framework of the CP-violating MSSM. We explored the $\ell jjjj \met$
channel with double or triple $b$ tag, focusing on the region of parameter
space not excluded by LEP searches. We used the PYTHIA event generator and
implemented $E_T$ dependent $b$ tagging and light-flavor mis-tagging on a
jet-by-jet basis. We found that this signal is impossible to observe at the
Tevatron, since it features quite soft $b$ jets, which have poor tagging
efficiencies. A preliminary study for the LHC shows survival of a few signal
events over a much larger background. We are extending this study without
employing $b$ tagging \cite{mdadmmspd} and also by using other Higgs decay
modes.


\begin{theacknowledgments}
  
SPD acknowledges financial support from the Bundesministerium f\"ur Bildung
und Forschung (BMBF) Projekt under Contract No. 05HT6PDA.

\end{theacknowledgments}

\end{document}